\documentclass[aps,twocolumn,preprintnumbers,amsmath,amssymb,floatfix]{revtex4}
\usepackage{amsmath,amssymb,latexsym}
\usepackage{tikz}
\usepackage{wrapfig}
\usepackage{times}
\usepackage{epsfig}
\usepackage{graphicx}
\usepackage{color}
\usepackage{dcolumn}
\usepackage{bm}
\usepackage{setspace}
\usepackage[center]{subfigure}
\usepackage{enumitem}
\usepackage{placeins}
\usepackage{pgfplots}
\usepackage{dcolumn}

\setlist[itemize]{noitemsep}

\newcommand{\sumlimits}[3]{\sum_{#2}^{#3}#1}
\newcommand{\intKernLeft}{\kern-.3em \int}
\newcommand{\intKernRight}{\int \kern-.3em}
\newcommand{\intKern}{\kern-.3em \int \kern-.3em}

\newcommand{\vect}[1]{\mathbf{#1}}
\newcommand{\unitvect}[1]{\vect{#1}}

\newcommand{\ft}[1]{\hat{#1}}
\newcommand{\primed}{^\prime}
\newcommand{\dprimed}{^{\prime \prime}}
\newcommand{\intd}{\kern.1em d}
\newcommand{\circFun}{\mathrm{circ}}

\begin{document}

\title{A New Structural Phase Field Crystal Approach for Modelling Graphene}
\author{Matthew Seymour$^1$}
\author{Nikolas Provatas$^1$}

\affiliation{$^1$ Department of Physics, Centre for the Physics of Materials, McGill University, 3600 Rue University, Montreal, QC, H3A2T8 Canada\\}

\date{\today}

\begin{abstract}
This paper introduces a new structural phase field crystal (PFC) type model that expands the PFC methodology to a wider class of structurally complex crystal structures than previously possible.
Specifically, our new approach allows for stabilization of graphene, as well as its coexistence with a disordered phase.
It also preserves the ability to model the usual triangular and square lattices previously reported in 2D PFC studies.
Our approach is guided by the formalism of classical field theory, wherein the the free energy functional is expanded to third order in  PFC density correlations.
It differs from previous PFC approaches in two main features.
First, it utilizes a hard-sphere repulsion to describe two-point correlations.
Second, and more important, is that it uses a rotationally invariant three-point correlation function that provides a unified way to control the formation of crystalline structures that can be described by a specific bond angle, such as graphene, triangular or square symmetries.
Our new approach retains much of the computational simplicity of previous PFC models and allows for efficient simulation of nucleation and growth of polycrystalline 2D materials. 
In preparation for future applications, this paper details the mathematical derivation of the model and its equilibrium properties, and uses dynamical simulations to demonstrate defect structures produced by the model. 
\end{abstract}

\maketitle

\section{Introduction}
Graphene is one of the most exciting new 2D material discovered. 
It has been found to exhibit interesting electrical \cite{Geim2007,Cai2009} and mechanical properties \cite{Stankovich2006,Lee2008}.
Crystals of graphene can be obtained by exfoliating cleaved graphite samples onto an oxidized silicon wafer to produce flakes of graphene \cite{Novoselev2004}.
A more scalable method of obtaining graphene is through CVD \cite{Li2011,Tetlow2014}, a method that produces a polycrystalline material.
While theoretically graphene can be about a hundred times stronger than steel, the properties of graphene realized in experiments typically reveal a wide variability, up to an order of magnitude from their  theoretical predictions \cite{Ruiz2011,Lee2013}. 

Variability in the strength properties of graphene, particularly how these are related to the defect structure, still remains largely unexplored, particularly theoretically.
Recent work suggests that it is linked to the defect microstructure at grain boundaries  \cite{Ruiz2011,Rasool2013,Lee2013}.
The topological defects in graphene typically take the form of periodic  patterns of heptagonal and pentagonal disclinations, the patterning of which are dictated by the tessellation requirements of atoms in adjacent grains \cite{Yazyev2010,Wei2012,Daly2014}. 
The complexity of forming and measuring graphene, however, make it challenging to experimentally isolate and examine the role of specific defects and grain boundaries on the growth and properties of this material. 

Computational modelling can serve as a route for theoretically understanding the difficult to measure properties of graphene. 
First principles studies are useful in examining the adsorption process of carbon onto metal surfaces during graphene formation \cite{Galea2007}.
Molecular dynamics (MD) studies of graphene have been successful at predicting the anisotropy of graphene morphologies on metal surfaces \cite{Vasili2015} or the anergy of specific defect structures \cite{Daly2014}.
On the continuum scale,  phase field models have been used to study how anisotropic diffusion of carbon on a surface can yield the formation of the dendritic graphene structures \cite{Meca2013}. 
To date, there has not been a model that can address both the atomically varying defect microstructures of graphene alongside its nucleation and diffusional growth kinetics from a disordered state on a surface.   

The phase field crystal (PFC) modelling approach is a promising approach for modelling many microstructure phenomena.
The approach describes the thermodynamics and dynamics of phase transformations through an atomically varying order parameter field that is loosely connected to the atomic density field.
Like traditional phase field (PF) models, PFC models naturally capture most of the salient physics of nucleation, polycrystalline solidification, grain boundaries \cite{ElderGrant,PlappMethod,GreenwoodPure,Toth2011,Granasy2014} and multi-component, multi-phase solidification \cite{ElderCDFT,Ofori-Opoku13,Ofori_Opoku13B,Kocher2015}.
Unlike traditional PF models, PFC models also capture, in the context of a single order parameter, elasticity and plasticity phenomena relevant to solid state processes such as dislocation source creation, dislocation stability \cite{Berry12,Berry14} and creep \cite{Berry15}.  
The most important feature of PFC-type models is that they incorporate the above phenomena from atomic to micron length scales and over diffusional times scales, where the emergent properties of non-equilibrium phase transformations are typically manifested.
PFC modelling has been used to elucidate phenomena ranging from grain boundary pre-melting  \cite{Berry08,Spatschek2010,PlappMethod} to the nucleation pathways of defect-mediated nucleation of precipitates, the latter of which led to TEM and atomic probe experiments to validate the PFC predictions \cite{Vahid2013}.

The original PFC model was predominately used for the study 2D triangular and 3D BCC crystal symmetries \cite{Elder,ElderGrant}.
Later models introduced multi-peaked two-point correlation kernels in the non-local part of the free energy that allowed for a simple yet robust manner to simulate most of the common metallic crystal structures (2D Square, BCC,FCC,HCP) in phase transformations \cite{GreenwoodPRL,Wu10b}.
These so-called structural PFC ({\it XPFC}) models were later generalized to binary and multi-component (and multi-phase) alloys \cite{GreenwoodAlloy,Ofori-Opoku13}.
More recently, a new multi-peaked two-point correlation was introduced to stabilize graphene and Kagome lattices, and a morphological phase diagram distinguishing the stability ranges of the two solid phases was explored numerically \cite{Mkhonta2013}. 

This paper introduces a new structural PFC theory that breaks with the tradition of previous PFC models and expands the free energy up to three--point correlations in the PFC density field. 
Unlike previous PFC theories, the two-point excess term is based on hard sphere-like interactions.
It is shown that this allows for stabilization of triangular symmetry in two dimensions. 
In this formalism,  more complex crystal structures that are describable by a particular bond angle are stabilized using a new, rotationally invariant, three-point correlation function we introduce for the excess free energy.
This term allows for the stabilization of triangular, square and graphene lattices in two dimensions. 
It is noteworthy that beyond stabilizing the aforementioned structures, this formalism also allows for stable coexistence of these structures with a disordered phase, a feature crucial for modelling nucleation and growth of polycrystalline 2D materials from a vapour or a disordered arrangement of atoms on a surface.  
In preparation for future applications, this paper highlights the derivation of our PFC free energy, examines its equilibrium properties and use dynamics to demonstrate defect structures produced in polycrystalline samples.

\section{New Structural PFC Model}
The derivation of our model begins by defining the spatial PFC density field, $\rho$, of a species of atoms.  
From this, a dimensionless density field is defined as $n=(\rho-\bar{\rho})/\bar{\rho}$, where $\bar{\rho}$ is the reference density of a disordered phase around which a functional expansion of the free energy is carried out.
Treating the field $n$ as an order parameter with which to describe microstructure variations, we expand the free energy of a crystallizing system as
\begin{equation}
\frac{\Delta F}{k_BT\bar{\rho}} = F_{id}[n] +F_{ex,2}[n]+ F_{ex,3}[n]
\label{newXPFC}
\end{equation}
The term $F_{id}$ is the ideal free energy, which ignores interactions.
Its form here is given by
\begin{align}
F_{id}=  \int d\vect{r} \, \left\{ \frac{n^2}{2} \!-\! \eta\frac{n^3}{6} \!+\! \chi\frac{n^4}{12} \right\} 
\label{newXPFC2}
\end{align}
where $\eta$ and $\chi$ are dimensionless parameters to adjust the form of the ideal free energy.
This form is a Landau expansion of the true ideal free energy.
In what follows $\eta = \chi = 1$. 
The term $F_{ex,2}$ is the first term in the expansion of excess free energy, which incorporates two-point interactions. 
Its form is written as 
\begin{align}
F_{ex,2} = -\frac{1}{2} \intKernRight n(\vect{r}) \intKern C_2 (\vect{r} - \vect{r}\primed) n(\vect{r}\primed) \intd \vect{r}\primed \intd \vect{r} \label{eq:two_point_excess}
\end{align}
where $C_2$ is the two-point correlation function. 
The term $F_{ex,3}[n]$ is the second term in the expansion of the excess free energy and describes interactions at the level of three-point correlations in the density $n$. 
Its form, assuming translational invariance is given by 
\begin{align}
F_{ex,3} \! =\! -\frac{1}{3} \! \intKernRight n(\vect{r}) \intKern C_3 (\vect{r} \!-\! \vect{r}\primed, \vect{r} \!-\! \vect{r}\dprimed) n(\vect{r}\primed) n(\vect{r}\dprimed) \intd \vect{r}\primed \! \intd \vect{r}\dprimed \! \intd \vect{r}
 \label{eq:three_point_excess}
\end{align}
where $C_3$ is the three-point correlation function. 
The remainder of this section will discuss the explicit forms of $C_2$ and $C_3$, and their mathematical properties.

\subsection{Two-point correlations}
\begin{figure}[h]
\centering
\includegraphics{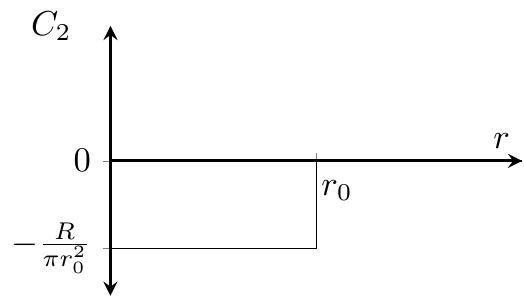}
\caption{
Two-point correlation function in real space.
\label{two_point_schematic}
}
\label{fig:two_point_correlation}
\end{figure}

We define $C_2$ in $F_{ex,2}$ using a simple repulsive term.
The length scale defined by this term will set the crystal lattice spacing. 
We define the $\circFun$ function as a circular step function of unit radius and height centred on the origin,
\begin{align}
\circFun(r) = \left\{
     \begin{array}{lr}
       1 & : r \le 1\\
       0 & : r > 1
     \end{array}
   \right.
\end{align}
In two dimensions we then define $C_2$ as
\begin{align}
C_2(\vect{r}) = - \frac{R}{\pi r_0^2} \circFun \left( \frac{r}{r_0} \right)
\label{eq:two_point_2d}
\end{align}
where $r_0$ sets the cutoff for the repulsive term and $R$ sets the magnitude of the repulsion. 
The normalization factor has been set such that in reciprocal space $\ft{C}_2(0) = -R$, where $\hat{C}_2(\vect{k})$ denotes the Fourier transform of $C_2(\vect{r})$. 
The form of $C_2(\vect{r})$ is depicted schematically in Fig.~\ref{fig:two_point_correlation}.

\begin{figure}[h]
\centering
\includegraphics{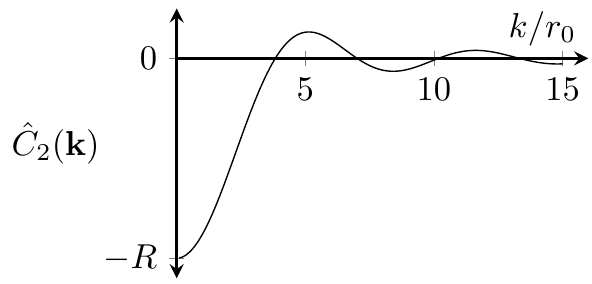}
\caption{
Plot of $- 2 R {J_1(r_0 k)}/{(r_0 k)}$ in units of $k/r_0$.
}
\label{fig:two_point_correlation_ft}
\end{figure}

It is convenient to simulate PFC models numerically in Fourier space. 
The Fourier transform of Eq.~(\ref{eq:two_point_2d}) becomes
\begin{align}
\ft{C}_2 (\vect{k}) = - 2 R \frac{J_1(r_0 k)}{r_0 k}
\label{C2infft}
\end{align}
where $J_m$ are the Bessel functions of the first kind.
Figure~\ref{fig:two_point_correlation_ft} shows a plot of $\ft{C}_2 (\vect{k})$.
Using Eq.~(\ref{C2infft}), the excess free energy due to two-point correlations in Eq.~(\ref{eq:two_point_excess}) can be written with the use of the convolution theorem as 
\begin{align}
F_{ex,2} = -\frac{1}{2} \intKernRight n(\vect{r}) \mathcal F^{-1} \big\{ \ft{C}_2 (\vect{k}) \ft{n}(\vect{k}) \big\} \intd \vect{r}
\end{align}

While there is no attraction between atoms at the level of two-point correlations, the system can still undergo a phase transition and solidify when the density is great enough. 
The lattice constant $a_0$ is related in a nontrivial way to the cutoff length $r_0$. 
Assuming for simplicity that most of the energy is carried in the first reciprocal space mode ($K_1$) of the density, the energy of the system will be minimized when the first mode lies on the peak of $- 2 R {J_1(r_0 k)}/{(r_0 k)}$, i.e., when $K_1 \approx 5.13562 / r_0$.
Table \ref{tb:R0_ratios} gives the ratio $r_0/a_0$ for various two dimensional lattices. 
\begin{table}
\begin{ruledtabular}
\begin{tabular}{lcd}
Lattice & $K_1 a_0$ & \multicolumn{1}{c}{$r_0 / a_0$} \\ 
\colrule
\vspace{-1em}  $ $ \\
Triangular & $4 \pi / \sqrt{3} $ & 0.707854 \ldots \\
Square        & $2 \pi$               & 0.81736 \ldots \\ 
Graphene   & $4 \pi / 3$            & 1.22604 \ldots \\
\end{tabular}
\end{ruledtabular}
\caption{
The ratio of $r_0$ to $a_0$ for various two dimensional crystal lattices. 
$K_1$ is the reciprocal lattice vector of the first mode of a crystal structure.
}
\label{tb:R0_ratios}
\end{table}

Since the two-point correlation function has only one equilibrium distance and is completely isotropic it strongly favours a lattice with the highest packing fraction.
In two dimensions this is the triangle phase, a result consistent with the classical result of hard sphere theory \cite{Kalikmanov}. 
To stabilize more structurally complex solid phases we must include either additional distances (as in XPFC models) or introduce a term which breaks the isotropy in interactions. 
Since we require our free energy to be rotationally invariant, this breaking of isotropy can only be relative to some local density configuration. 
This is not possible to achieve with a two-point correlation function -- we must proceed to higher order and consider three-point correlation functions. 
This will permit us to favour particular relative angles between nearest neighbour atoms in order to produce crystal structures of interest, particularly those of non-metals.

\subsection{Three-point correlations}
Three-point correlations in the model are described by the excess energy in Eq.~(\ref{eq:three_point_excess}). 
This term is computationally expensive compared to the two-point excess term of Eq.~(\ref{eq:two_point_excess}). 
By use of the convolution theorem the two-point correlation can be computed by transforming to reciprocal space and multiplying point-wise. 
The computational complexity is therefore on the order of $\mathcal O( N \log N)$, the complexity of the fast Fourier transform (where $N$ is the total number of grid points in the system). 
There is no such reduction in complexity available for Eq.~(\ref{eq:three_point_excess}) to our knowledge. 
By comparison, it requires a prohibitive $\mathcal O( N^3)$ number of calculations, making it impractical for most purposes. 

To remedy this problem we propose to separate $C_3 (\vect{r}-\vect{r}', \vect{r}-\vect{r}^{''})$ in the following manner
\begin{align}
C_3 (\vect{r}-\vect{r}', \vect{r}-\vect{r}'') = \sumlimits{C_s^{(i)} (\vect{r}-\vect{r}') C_s^{(i)} (\vect{r}-\vect{r}'')         }{i}{}
\label{new_3p}
\end{align}
where the $C_s^{(i)}$ will be defined below.
While this separation limits the possible forms of $C_3$, it is sufficiently flexible to produce a wide variety of crystal structures, including the ones previously modelled by XPFC and similar 2D models.  
When Eq.~(\ref{new_3p}) is inserted into Eq.~(\ref{eq:three_point_excess}) we obtain
\begin{align}
F_{ex,3} \!=\! -\frac{1}{3} \! \intKernRight n(\vect{r}) \sumlimits{ \left( \intKernRight C_s^{(i)} (\vect{r} - \vect{r}\primed) n(\vect{r}\primed)  \intd \vect{r}\primed  \right)^2    }{i}{} \intd \vect{r} \label{eq:three_point_excess_sep}
\end{align}
Details on how to approach this term computationally will be discussed below.
Here it should be apparent that in reducing the integration over $\vect{r}\primed$ and $\vect{r}\dprimed$ to one just over $\vect{r}\primed$ considerably reduces the computational complexity of the three-point  term. 

We next define the $C_s^{(i)}$ functions. 
We work in polar coordinates, and separate  $\vect{r}$ into $r$ and $\theta$ according to
\begin{gather}
C_s^{(1)}(r, \theta) = C_r (r) C_{\theta}^{(1)}(\theta) = C_r (r) \cos( m \theta) \label{eq:Ctheta1}\\
C_s^{(2)}(r, \theta) = C_r (r) C_{\theta}^{(2)}(\theta) = C_r (r) \sin( m \theta) \label{eq:Ctheta2}\\
C_r (r) = \frac{X}{2 \pi a_0} \delta(r - a_0) \label{Cs}
\end{gather}
Where $X$ is a parameter defining the strength of the interaction, $a_0$ corresponds to the lattice spacing, and $m$ defines bond order (discussed below) of the crystal phase.  

It may appear as though Eqs.~(\ref{eq:Ctheta1})-(\ref{eq:Ctheta2}) break the isotropy of the free energy, but in fact they do not. 
While each term in the sum over $i$ in Eq.~(\ref{eq:three_point_excess_sep}) exhibits angular dependence, the total sum is rotationally invariant. 
To see this, consider the sum in Eq.~(\ref{eq:three_point_excess_sep}) and expand the square. 
This gives
\begin{align}
& \sumlimits{ \left( \intKernRight C_s^{(i)} (\vect{r} - \vect{r}\primed) C_s^{(i)} (\vect{r} - \vect{r}\dprimed) n(\vect{r}\primed)  n(\vect{r}\dprimed) \intd \vect{r}\primed \intd \vect{r}\dprimed  \right)  }{i}{} 
\end{align}
Defining $\vect{r_1} \equiv  \vect{r} - \vect{r}\primed$, $\vect{r_2} \equiv  \vect{r} - \vect{r}\dprimed$ and pulling the sum inside the integral gives
\begin{align}
= &\intKernRight \left( \sumlimits{ C_s^{(i)} (\vect{r_1}) C_s^{(i)} (\vect{r_2})}{i}{} \right) n(\vect{r} - \vect{r_1})  n(\vect{r} - \vect{r_2}) \intd \vect{r_1} \intd \vect{r_2} 
\end{align}
Considering the sum in the brackets, changing to polar coordinates $\vect{r_i} \rightarrow (r_i, \theta_i$)  gives
\begin{gather}
 \sumlimits{ C_s^{(i)} (\vect{r_1}) C_s^{(i)} (\vect{r_2})}{i}{} =  \sumlimits{ C_s^{(i)} (r_1,\theta_1) C_s^{(i)} (r_2,\theta_2)}{i}{}\\
 = C_r(r_1) C_r(r_2) \sumlimits{ C_{\theta}^{(i)} (\theta_1) C_{\theta}^{(i)} (\theta_2)}{i}{}
 \end{gather}
Inserting the $C_{\theta}^{(i)}$ from Eqs.~(\ref{eq:Ctheta1}) and (\ref{eq:Ctheta2}) yields 
\begin{multline}
\sumlimits{ C_s^{(i)} (\vect{r_1}) C_s^{(i)} (\vect{r_2})}{i}{} = \\ C_r(r_1) C_r(r_2) \Big\{ \cos(m \theta_1) \cos(m \theta_2)  \\ + \sin(m \theta_1) \sin(m \theta_2) \Big\}
\end{multline}
Finally, noting the identity 
\begin{align}
\cos (\theta_1) \cos (\theta_2) + \sin (\theta_1) \sin (\theta_2) = \cos (\theta_2 - \theta_1)
\end{align}
gives 
\begin{align}
\!\!\! \! \sumlimits{ \!C_s^{(i)} (\vect{r_1}) C_s^{(i)} (\vect{r_2})}{i}{} \!=\! C_r(r_1) C_r(r_2) \cos \big(m (\theta_2 - \theta_1) \big)
\label{eq:three_point_excess_angle}
\end{align}
Since Eq.~(\ref{eq:three_point_excess_angle}) depends only on the difference $\theta_2 - \theta_1$, the free energy remains isotropic. 
Moreover, by selecting the appropriate values for $m$ we can favour certain crystal bond angles over others. 
For example, $m = 6$ favours six-fold triangular crystals, $m = 4$ favours four-fold square crystals and $m = 3$ favours three-fold graphene crystals.

It will be useful below to have the Fourier transforms of $C_s^{(i)}(r, \theta)$. 
Starting with $C_s^{(1)}(r, \theta)$, we rewrite it as a multipole expansion, 
\begin{align}
C_s^{(1)}(r, \theta) &= \frac{X}{2\pi a_0} \delta(r-a_0) \frac{ e^{i m \theta} + e^{-i m \theta} }{2} 
\end{align}
which transforms as
\begin{align}
\ft{C}_s^{(1)}(k, \theta_k) &=X \frac{ i^m e^{i m \theta_k} J_m (k a_0) + i^{-m} e^{-i m \theta_k} J_{-m} (k a_0)}{2} \\
                                        &=X i^m \frac {e^{i m \theta_k} + e^{-i m \theta_k} }{2}  J_m (k a_0) \label{eq:csfft} \\
                                        &= X i^m \cos(m \theta_k) J_m (k a_0)
\end{align}
where ($k$, $\theta_k$) are the polar coordinates in Fourier space. 
In Eq.~(\ref{eq:csfft}) we have used the fact that $J_{-m}(r) = (-1)^m J_{m}(r)$. 
Proceeding similarly for  $C_s^{(2)}(r, \theta)$ gives  
\begin{align}
\ft{C}_s^{(2)}(k, \theta_k) &= X i^m \sin(m \theta_k) J_m (k a_0)
\end{align}

\section{Equilibrium Properties of Model}
This section derives the equilibrium properties of our new structural PFC model defined by Eqs.~(\ref{newXPFC})-(\ref{eq:two_point_excess}) and (\ref{eq:three_point_excess_sep}) with Eqs.~(\ref{eq:Ctheta1})-(\ref{Cs}). 
In particular, we construct phase diagrams describing solid-disorder coexistence for the case of three crystal systems; graphene, triangular and square. 

\begin{figure}[h]
\centering
\includegraphics{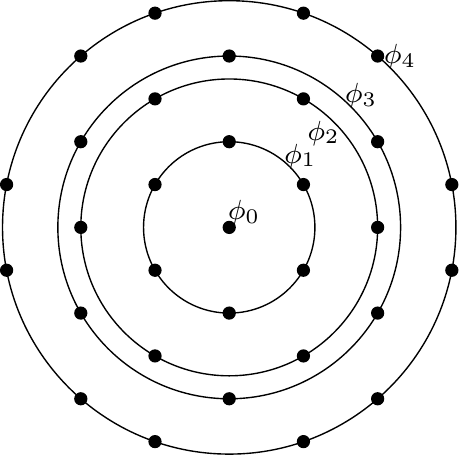}
\caption{
Organization of amplitudes according to modes. 
Dots represent reciprocal lattice vectors, circles connect vectors of equal magnitude, i.e. in the same mode.
}
\label{fig:mode}
\end{figure}

To describe the equilibrium properties of the model, we can expand the density of a crystal structure in a Fourier series:
\begin{align}
n(\vect{r}) &= \sumlimits{\phi_{\vect{q}} e^{i \vect{q} \cdot \vect{r}} }  {\vect{q}}{},
\label{eq:crystal_fourier}
\end{align} 
where $\vect{q}$ are the reciprocal lattice vectors of the crystal structure. 
The vectors $\vect{q}$ can be grouped according to their magnitude; these groups are collectively referred to as \textit{modes}, which are indexed by the integer $k$.
The vectors within a mode are indexed by $j$, and the notation $\vect{q}_{k,j}$ is used to refer to a vector $j$ of mode $k$.
To simplify the equilibrium analysis, we assume that all amplitudes $\phi_{\vect{q}}$ of a given mode are equal and real, i.e., $\phi_{\vect{q}_{k,j}} = \phi_{k}$ $\forall ~j$ (see Figure~\ref{fig:mode}).
With these changes our expansion becomes
\begin{align}
n(\vect{r}) &\approx \sumlimits{\phi_{k}  \left( \sumlimits{e^{i \vect{q}_{k,j} \cdot \vect{r}}} {j}{} }  {k=0}{N} \right),
\label{eq:mode_approx}
\end{align} 
where we have truncated the expansion to $N$ modes.  

To determine equilibrium properties of the model we use an approach well documented in numerous other PFC papers papers \cite{ElderCDFT}.
We begin by inserting Eq.~(\ref{eq:mode_approx}) into the ideal and excess free energies and integrating over a unit cell.
Since $\phi_k$ are constants, the integration removes the dependency on $\vect{r}$, leaving a free energy dependent on the amplitudes $\phi_k$ and the model parameters ($R$ and $X$).
The constant $\phi_0$ corresponds to the average system density, while $\phi_k$, $k>0$, comprise a set of order parameters for the crystal phase. 
Minimizing the free energy numerically with respect to the $\phi_k$ determines the phase and free energy of the system for a given $\{\phi_0, R, X\}$. 
When $\phi_k = 0$ the system is in the disordered state. 
For non-zero $\phi_k$ the system is in a crystal state.
Mapping out the convex hull of the free energy as a function of $\phi_0$ gives a common tangent line that defines coexistence values of $\phi_0$ corresponding to the solid and disordered phases. 

\subsection{Triangular-disorder coexistence}
\label{HCP_3}
\begin{figure}[h]
\centering
\includegraphics{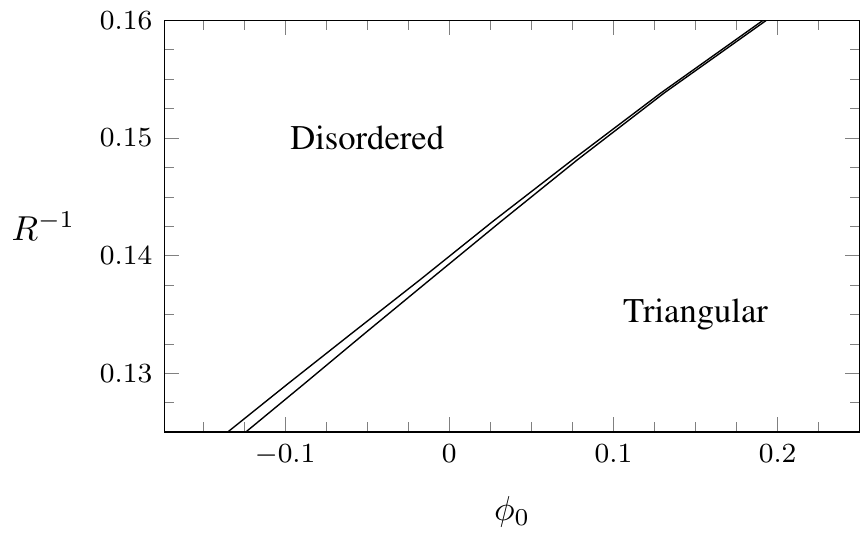}
\caption{
Triangular-disorder phase diagram using only two-point correlations ($X=0$).
}
\label{fig:phase_diagram_tri_no_X}
\end{figure}

Using the above procedure with a five mode expansion for a 2D triangular phase, we produce a phase diagram for triangular-disorder coexistence.  
As noted above, the rejection term $R$ is sufficient to produce a triangular phase without the need for three-point correlations. 
Figure~\ref{fig:phase_diagram_tri_no_X} shows the phase diagram in $\{\phi_0, R\}$ space obtained by setting $X=0$.

\begin{figure}[h]
\centering
\includegraphics{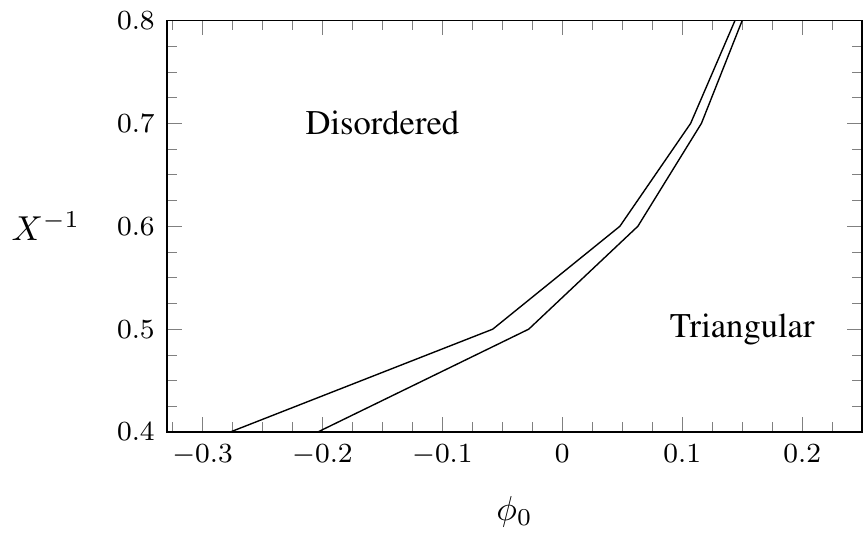}
\caption{
Triangular-disorder phase diagram using two and three-point correlations, with $m=6$ and $R=6$. 
Here $r_0/a_0=0.70785$.
}
\label{fig:phase_diagram_tri_X}
\end{figure}

With $R$ fixed we can also produce a triangular-disorder phase diagram in $\{\phi_0, X\}$ space.
Setting the three-point correlation function set to produce six-fold symmetry ($m=6$), Figure~\ref{fig:phase_diagram_tri_X} shows such a phase diagram with $R=6$. 

\subsection{Square-disorder coexistence}
\begin{figure}[h]
\centering
\includegraphics{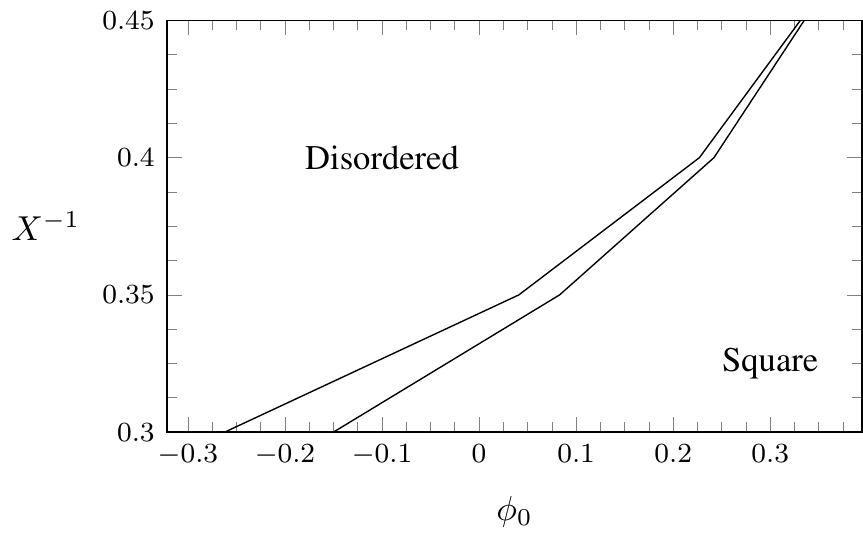}
\caption{
Square-disorder phase diagram using two and three-point correlations with $m = 4$ and $R = 5$.
Here $r_0/a_0=0.81736$.
}
\label{fig:phase_diagram_sq}
\end{figure}

To simulate square phases we require both two and three-point correlation functions in the free energy. 
To calculate the phase diagram for square-disorder coexistence, we use a five mode density expansion to describe the square phase, and set $m=4$. 
Figure~\ref{fig:phase_diagram_sq} shows a square-disorder phase diagram constructed with $R=5$. 

\subsection{Graphene-disorder coexistence}

The graphene crystal structure can be described using a triangular lattice with a two atom basis.
The primitive vectors of the triangular lattice are
\begin{align}
\vect{a_0} = \sqrt{3} \unitvect{x}; \qquad \vect{a_1} = \frac{\sqrt{3}}{2} \unitvect{x} + \frac{3}{2} \unitvect{y},
\end{align}
while the basis atoms are located at
\begin{align}
\vect{d}_0 = -\frac{\sqrt{3}}{2} \unitvect{x} - \frac{1}{2} \unitvect{y}; \qquad \vect{d}_1 = -\frac{\sqrt{3}}{2} \unitvect{x} + \frac{1}{2} \unitvect{y} \label{eq:basis_graphene}
\end{align}
We must distinguish between graphene and triangular phases on the basis of the amplitudes $\phi_{k}$ of the density expansion.
Assume we have a density expansion of a triangular phase $n_{\mathrm{H}}$ as in Eq~(\ref{eq:mode_approx}).
From this we can construct a graphene density field by 
\begin{align}
n_{\mathrm{G}}(\vect{r}) = \sumlimits{n_{\mathrm{H}}(\vect{r} - \vect{d}_i)}{i}{}
\end{align}
This gives
\begin{align}
n_{\mathrm{G}}(\vect{r}) &= \sumlimits{\phi_{k}  \left( \sumlimits{S_{k,j} e^{i \vect{q}_{k,j} \cdot \vect{r}}} {j}{} }  {k=0}{N} \right); \\
S_{k,j} &= \sumlimits{e^{-i \vect{q}_{k,j} \cdot \vect{d_i}}}{i}{},
\end{align}
where $S_{k,j}$ are the structure factors for graphene.
The basis vectors in Eq.~(\ref{eq:basis_graphene}) have been selected such that the structure factors are independent of $j$, and so for our purposes $S_{k,j} = S_k$.
Computing the first four structure factors we find
\begin{align}
S_0 = 2; \quad S_1 = -1; \quad S_2 = 2; \quad S_3 = -1; \label{eq:graphene_structure_factors}
\end{align}
Note that some modes have negative structure factors.
This is in contrast to the structure factors of the triangular crystal structure, which are all unity. 
We can thus distinguish between the triangular and graphene phases by the signs of $\phi_k$: when $\phi_1 < 0$ we have a graphene phase and where $\phi_1 > 0$ we have a triangular phase.

The presence of these negative amplitudes for graphene make it difficult to stabilize graphene-like structures using a two-point correlation function alone. 
This can be seen by inserting the amplitude expansion for a triangular lattice into the two-point correlation term in Eq.~(\ref{eq:two_point_excess}) and integrating over a unit cell, giving 
\begin{align}
F_{ex,2} = -\frac{1}{2} \left(\ft{C}_2 (q_0)  \phi_0^2 + 6 \ft{C}_2 (q_1) \phi_1^2 + 6 \ft{C}_2 (q_2) \phi_2^2 + \ldots \right)
\label{eq:exexpansion}
\end{align}
Since all amplitudes in Eq.~(\ref{eq:exexpansion}) are squared, the two-point correlation function alone cannot break the symmetry between positive and negative amplitudes, the latter of which are required to stabilize graphene. 
Moreover, since the ideal free energy is minimized by positive amplitudes, the triangular phase results.
The three-point correlation function, however, results in terms of odd power which make it possible to minimize the free energy with negative amplitudes in order to produce a graphene phase.

\begin{figure}[h]
\centering

\includegraphics{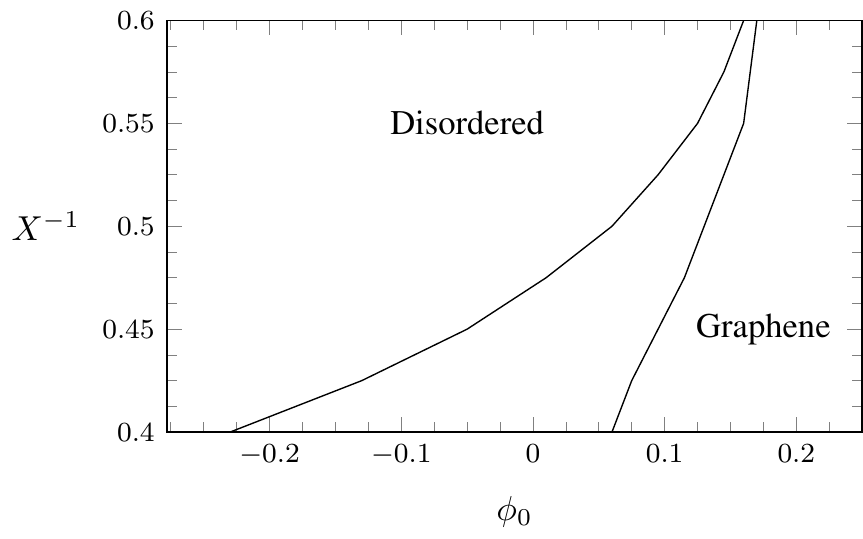}
\caption{
Graphene-disorder coexistence phase diagram, with $m=3$ and $R=6$.
Here $r_0/a_0 = 1.2259$.
}
\label{fig:phase_diagram}
\end{figure}

Using a five mode triangular density expansion in the free energy with $m=3$ and applying the common tangent construction yields a graphene-disorder coexistence phase diagram in $\{\phi_0, X\}$ space.
Figure~\ref{fig:phase_diagram} shows such a phase diagram for the case $R=6$.
As expected, it was found that $\phi_1 < 0$ for the ordered region of the phase diagram.

\section{Dynamical Simulations}
In this section we demonstrate the dynamical stability of the phase coexistence predicted by the phase diagrams of the previous section.
We also demonstrate the robustness of our model to simulate nucleation, growth and formation or polycrystalline graphene, square and triangular crystal phases from a disordered phase.
For the case of graphene, we also demonstrate the formation of experimentally relevant of defect structures at grain boundaries.

The density $n$ represents a conserved order parameter. 
As a result, its evolution is described by model B dynamics for conserved fields \cite{Hoh1977}. 
This gives
\begin{align}
\frac{\partial n}{\partial t} = M_n \nabla^2 \left( \frac{\delta F}{\delta n} \right)
\label{modelb}
\end{align}
Where $M$ is an effective mobility that sets the scale of the diffusional dynamics of $n$. 
Equation~(\ref{modelb}) should also have a noise source to subsume the role of thermal fluctuations. 
In this work noise will be ignored.  

To simulate Eq.~(\ref{modelb}), it is instructive to compile the various terms in the variational $\delta F / \delta n$. 
The variational of the ideal term is straight forward. 
The variational of the two-point excess term in Eq.~(\ref{eq:two_point_excess}) becomes
\begin{align}
\frac{\delta F_{ex,2}}{\delta n} = -\intKern C_2 (\vect{r} - \vect{r}\primed) n(\vect{r}\primed) \intd \vect{r}\primed \equiv -C_2 \ast n
\end{align}
Where $\ast$ indicates convolution. 
Terms such as this may be efficiently computed in reciprocal space by use of the convolution theorem. 

The three-point excess term of Eq.~(\ref{eq:three_point_excess_sep}) is more complex. 
For each term in the summation we expand the square,
\begin{multline}
F_{ex,3}^{(i)}[n] = \\ 
-\frac{1}{3} \intKernRight C_s^{(i)} (\vect{r} - \vect{r}\primed) C_s^{(i)} (\vect{r} - \vect{r}\dprimed) n(\vect{r}) n(\vect{r}\primed) n(\vect{r}\dprimed)  \intd \vect{r}\primed \intd \vect{r}\dprimed    \intd \vect{r}
\end{multline}
We find the functional derivative using the well known formula \cite{Chaikin1995}
\begin{align}
F[n + \delta n] - F[n] \equiv \intKern \delta n \frac{\delta F}{\delta n} \intd \vect{r}
\end{align}
Applying this to Eq.~(\ref{eq:three_point_excess_sep}) and discarding terms of order $\mathcal O \big( (\delta n)^2 \big)$ we are left with three terms,
\begin{align}
& F_{ex,3}^{(i)}[n + \delta n] - F_{ex,3}^{(i)}[n] = \nonumber \\
& -\frac{1}{3} \intKernRight C_s^{(i)} (\vect{r} - \vect{r}\primed) C_s^{(i)} (\vect{r} - \vect{r}\dprimed) \delta n(\vect{r}) n(\vect{r}\primed) n(\vect{r}\dprimed)  \intd \vect{r}\primed \intd \vect{r}\dprimed    \intd \vect{r} \nonumber \\
& -\frac{1}{3} \intKernRight C_s^{(i)} (\vect{r} - \vect{r}\primed) C_s^{(i)} (\vect{r} - \vect{r}\dprimed)  n(\vect{r})  \delta n(\vect{r}\primed) n(\vect{r}\dprimed)  \intd \vect{r}\primed \intd \vect{r}\dprimed    \intd \vect{r}  \nonumber  \\
& -\frac{1}{3} \intKernRight C_s^{(i)} (\vect{r} - \vect{r}\primed) C_s^{(i)} (\vect{r} - \vect{r}\dprimed)  n(\vect{r}) n(\vect{r}\primed) \delta n(\vect{r}\dprimed)  \intd \vect{r}\primed \intd \vect{r}\dprimed    \intd \vect{r}
\end{align}
Swapping $\vect{r}$ and $\vect{r}\primed$ in the second term and $\vect{r}$ and $\vect{r}\dprimed$ in the third gives:
\begin{align}
& F_{ex,3}^{(i)}[n + \delta n] - F_{ex,3}^{(i)}[n] = \nonumber \\
& \qquad -\frac{1}{3} \intKernRight \delta n(\vect{r}) \intKernLeft \Big( C_s^{(i)} (\vect{r} - \vect{r}\primed) C_s^{(i)} (\vect{r} - \vect{r}\dprimed)  n(\vect{r}\primed) n(\vect{r}\dprimed)   \nonumber \\
& \qquad + C_s^{(i)} (\vect{r}\primed - \vect{r}) C_s^{(i)} (\vect{r}\primed - \vect{r}\dprimed)  n(\vect{r}\primed) n(\vect{r}\dprimed)   \nonumber  \\
& \qquad + C_s^{(i)} (\vect{r}\dprimed  - \vect{r}\primed) C_s^{(i)} (\vect{r} \dprimed - \vect{r})  n(\vect{r}\primed)  n(\vect{r}\dprimed) \Big)  \intd \vect{r}\primed \intd \vect{r}\dprimed     \intd \vect{r}
\end{align}
Combining the last two terms (by swapping $\vect{r}\primed$ and $\vect{r}\dprimed$ in the third term) gives
\begin{align}
& F_{ex,3}^{(i)}[n + \delta n] - F_{ex,3}^{(i)}[n] = \nonumber \\
& \qquad -\frac{1}{3} \intKernRight \delta n(\vect{r}) \intKernLeft \Big( C_s^{(i)} (\vect{r} - \vect{r}\primed) C_s^{(i)} (\vect{r} - \vect{r}\dprimed)  n(\vect{r}\primed) n(\vect{r}\dprimed)   \nonumber \\
& \qquad + 2 C_s^{(i)} (\vect{r}\primed - \vect{r}) C_s^{(i)} (\vect{r}\primed   - \vect{r}\dprimed)  n(\vect{r}\primed)  n(\vect{r}\dprimed) \Big)  \intd \vect{r}\primed \intd \vect{r}\dprimed     \intd \vect{r}
\end{align}
Finally it is noted from Eqs.~(\ref{eq:Ctheta1}) and (\ref{eq:Ctheta2}) that $C_s^{(i)} (-\vect{r}) = (-1)^m C_s^{(i)} (\vect{r})$. 
This leads to
\begin{align}
\frac{\delta F_{ex,3}^{(i)}}{\delta n} = -\frac{1}{3} \Big( \big[C_s^{(i)} \ast n]^2 + 2 (-1)^m C_s^{(i)} \ast \big[ n  \cdot (C_s^{(i)} \ast n) \big] \Big)
\end{align}
The first term can be computed by performing the convolution in reciprocal space and then returning to real space in order to compute the square. 
The second term can be computed in several steps: performing the inner convolution in reciprocal space; returning to real space for the multiplication by $n$; and finally transforming to reciprocal space once again to compute the outer convolution.

\subsection{Polycrystalline 2D materials, defects and coexistence}
\begin{figure}
\centering
\includegraphics{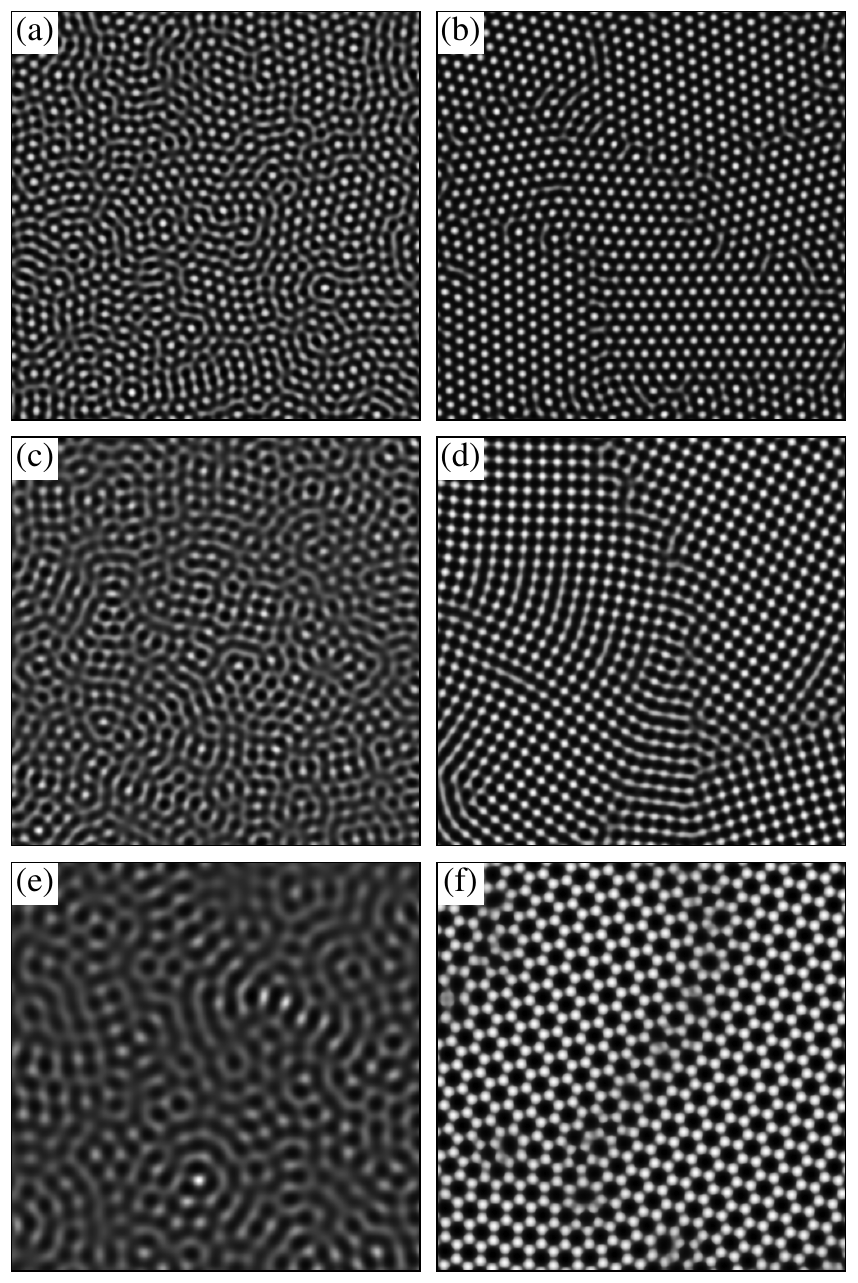}
\caption{
Density fields of triangular (\textit{a} and \textit{b}), square (\textit{c} and \textit{d}) and graphene (\textit{e} and \textit{f}) growth, showing early (left) and late (right) times during solidification.
Systems are initialized with gaussian density fluctuations.
For all three systems $\phi_0 = 0.3$ and values of $r_0 / a_0$ match those in Table~\ref{tb:R0_ratios}.
For triangular, $R = 7$ and $X=0$.
For square, $R = 6$ and $X^{-1}=0.5$.
For graphene, $R = 6$ and $X^{-1}=0.4$.
}
\label{fig:results}
\end{figure}

\begin{figure}
\centering
\includegraphics{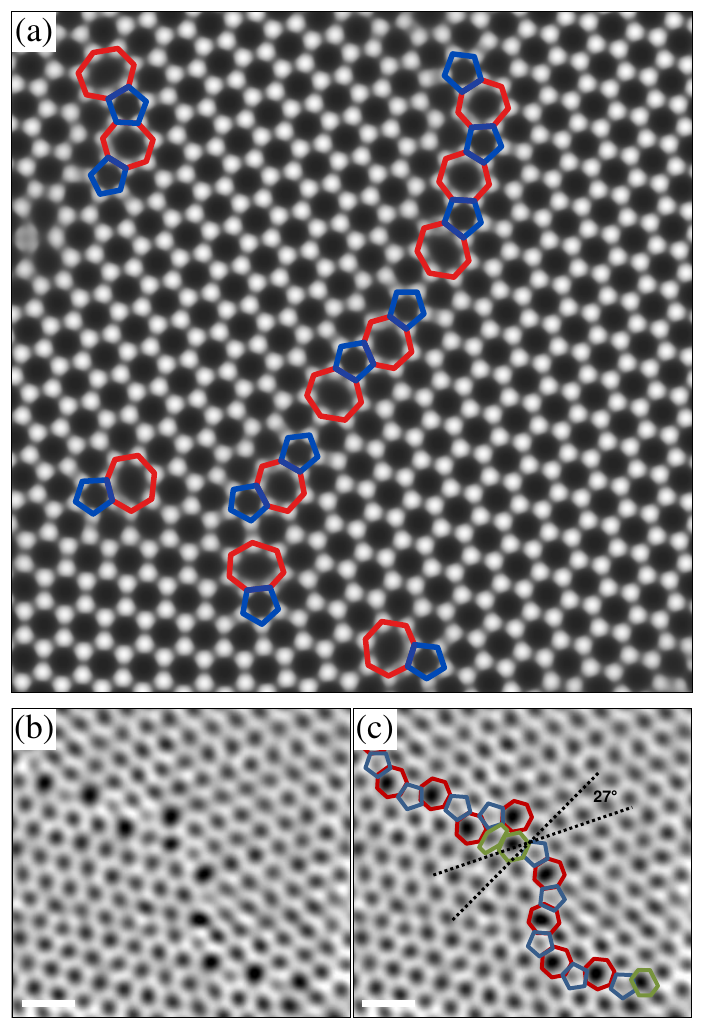}
\caption{
Comparison of simulated and experimentally determined defect structures of polycrystalline graphene.
The defect structure of Figure~\ref{fig:results}(\textit{f}) is highlighted in (\textit{a}).
The grain boundary is resolved by a line of 5-7 defect structures.
These defect structures match those found experimentally in polycrystalline graphene membranes grown by chemical vapour deposition (CVD) \cite{huang2011grains}.
(\textit{b}) shows an atomic resolution transmission electron microscope (TEM) image of one such graphene membrane; the defect structure is highlighted in (\textit{c}).
(\textit{b}) and (\textit{c}) reprinted by permission from Macmillan Publishers Ltd: Nature \cite{huang2011grains}, copyright 2011.
}
\label{fig:5-7_defects}
\end{figure}
We simulate the growth of triangular, square and graphene phases by choosing parameters corresponding to the solid region of the phase diagram and initializing the system with gaussian noise.
The system subsequently solidifies into a polycrystalline solid. 
Figure~\ref{fig:results} shows early and late time frames for the density of triangular, square and graphene systems under crystallization.

The defect structures which emerge along the graphene grain boundaries are noteworthy. 
Closer inspection of grain boundaries such as those in Figure~\ref{fig:results}(\textit{f}) reveals that where misaligned grains impinge the grain boundary is resolved into a line of so-called 5-7 (pentagons and heptagons) defects.
Figure~\ref{fig:5-7_defects} highlights these 5-7 defects.
These defects are in excellent agreement with the structures seen experimentally in polycrystalline graphene \cite{huang2011grains}. 
Like these experimentally determined structures, our simulated grain boundary consists of an aperiodic line of 5-7 defects.

\begin{figure}[h]
\centering
\includegraphics{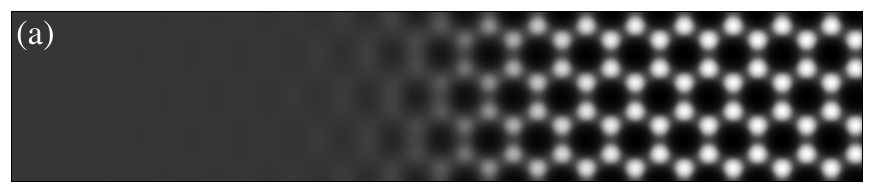}
\includegraphics{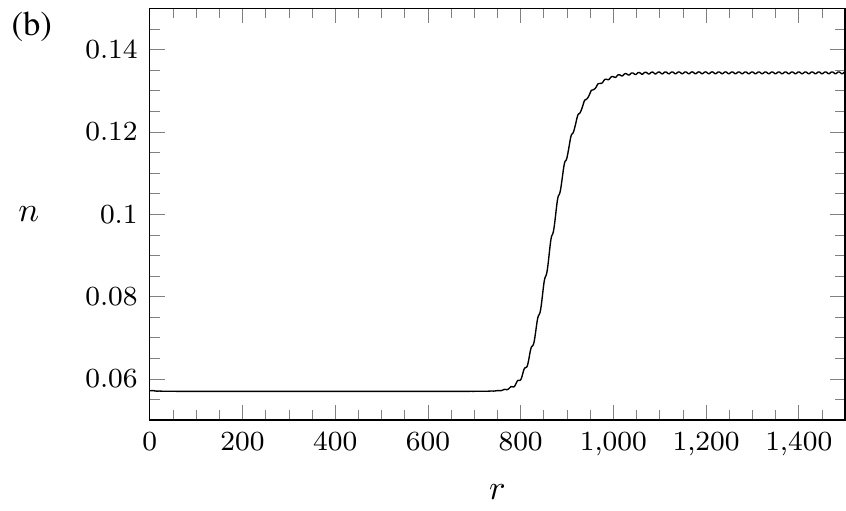}

\caption{
Simulation of coexistence between the ordered and disordered phases of graphene.
Density field $n(\vect{r}$) of the equilibrium interface between phases shown in (\textit{a}). 
Smoothed average density along the longitudinal axis depicted in (\textit{b}). 
$X^{-1}=0.5$, $R=6$. 
Average densities of $0.057$ and $0.134$ in the disordered and ordered phases respectively match closely the theoretical values from the phase diagram in Figure~\ref{fig:phase_diagram}.
\label{fig:coexistence}
}
\end{figure}

To demonstrate dynamic coexistence between the graphene and disordered phases, a $2000 \times 100$ system with periodic boundary conditions was seeded with a large initial slab of graphene and allowed to reach equilibrium with the disordered phase (Figure~\ref{fig:coexistence}(\textit{a})).
We set $X^{-1} = 0.5$ and $R = 6$.
This approach, where the slab extends through the system traverse to the long dimension, negates the effects of curvature on the equilibrium coexistence densities since the order-disorder interface is a straight line.
Figure~\ref{fig:coexistence}(\textit{b}) shows the smoothed density across the interface in the longitudinal direction. 
It can be seen that the equilibrium coexistence densities in the ordered and disordered regions closely match those of the phase diagram in Figure~\ref{fig:phase_diagram}.

\section{Conclusions}
This paper introduced the formalism of a new structural PFC theory that is truncated at three-point density correlations in the excess free energy.
This approach makes it possible to simulate microstructural evolution in metallic {\it and} non-metallic  materials, as well as their coexistence with a disordered phase.
Among the most important novel materials that can be studied with our new formalism is polycrystalline graphene. 

Our approach differers from previous ones in two major ways.
First of all, it treats two-point correlations more formally through the use of hard-sphere interactions.
As a result, the crystallography of structurally more complex phases than 2D triangular must be described in a unified way through the new rotationally invariant 3-point correlation introduced in this work. 
We showed that the form of our three-point correlation is rotationally invariant and robust enough to capture all crystal structures described through a single bond angle.  

After deriving the mathematical details of our new model, we calculated its equilibrium properties.
We then used dynamical simulations to illustrate the growth of polycrystalline graphene and other solids, and dynamical coexistence of graphene with a disordered phase.
We also compared the defect structures generated at grain boundaries against corresponding results from the experimental literature, finding excellent agreement. 

It is expected that the structural PFC formalism introduced here is easily amenable to the recent formalism of Ref.~\cite{Kocher2015}, whereby the addition of an additional long-wavelength interaction energy term can be used to bring the  pressure (P), $X$ and $\phi_0$ axes simultaneously under control.
Similarly, our model is extendable to multiple components.
These additions and their subsequent applications will be presented in future papers. 

\begin{acknowledgments}
The authors thank The National Science and Engineering Research Council of Canada and the Canada Research Chairs for funding, and Compute Canada for high performance computing resources. 
\end{acknowledgments}

\label{Bibliography}

\bibliography{Bibliography} 
\end{document}